\documentclass[prb,twocolumn,amsmath,amssymb,superscriptaddress]{revtex4}
\usepackage[dvips]{epsfig}
\usepackage{float}
\usepackage{amsmath}
\usepackage{amssymb}
\usepackage{amsfonts}
\usepackage{euscript}
\usepackage{enumerate}
\usepackage{hhline}
\usepackage{threeparttable}
\usepackage{pslatex}
\usepackage{tabularx}
\usepackage{color}

\usepackage{graphicx}
\usepackage{dcolumn}
\usepackage{bm}
\usepackage[sort&compress]{natbib}

\makeatletter
\renewcommand{\@biblabel}[1]{#1. }
\renewcommand{\@dotsep}{500}
\renewcommand{\@pnumwidth}{0em}
\renewcommand{\l@figure}[2]{
        \@dottedtocline{1}{1.5em}{2em}{Figure #1}{}\vspace{15pt}}

\begin{document}


\title{Quantum Transduction of Telecommunications-band Single Photons from a Quantum Dot by Frequency Upconversion}

\author{Matthew T. Rakher} \thanks{These authors contributed equally to this work.}
\affiliation{Center for Nanoscale Science and Technology, National
Institute of Standards and Technology, Gaithersburg, MD 20899, USA}
\author{Lijun Ma} \thanks{These authors contributed equally to this work.}
\affiliation{Information Technology Laboratory, National Institute
of Standards and Technology, Gaithersburg, MD 20899, USA}
\author{Oliver Slattery}
\affiliation{Information Technology Laboratory, National Institute
of Standards and Technology, Gaithersburg, MD 20899, USA}
\author{Xiao Tang} \email{xiao.tang@nist.gov}
\affiliation{Information Technology Laboratory, National Institute
of Standards and Technology, Gaithersburg, MD 20899, USA}
\author{Kartik Srinivasan} \email{kartik.srinivasan@nist.gov}
\affiliation{Center for Nanoscale Science and Technology, National
Institute of Standards and Technology, Gaithersburg, MD 20899, USA}

\date{\today}

\begin{abstract}
\textbf{The ability to transduce non-classical states of light from
one wavelength to another is a requirement for integrating disparate
quantum systems that take advantage of telecommunications-band
photons for optical fiber transmission of quantum information and
near-visible, stationary systems for manipulation and storage. In
addition, transducing a single-photon source at 1.3 $\mu$m to
visible wavelengths for detection would be integral to linear
optical quantum computation due to the challenges of detection in
the near-infrared. Recently, transduction at single-photon power
levels has been accomplished through frequency upconversion, but it
has yet to be demonstrated for a true single-photon source. Here, we
transduce the triggered single-photon emission of a semiconductor
quantum dot at 1.3 $\mu$m to 710 nm with a total detection (internal
conversion) efficiency of 21~$\%$ (75~$\%$). We demonstrate that the
710 nm signal maintains the quantum character of the 1.3 $\mu$m
signal, yielding a photon anti-bunched second-order intensity
correlation, $g^{(2)}(\tau)$, that shows the optical field is
composed of single photons with $g^{(2)}(0)=0.165<0.5$.  }
\end{abstract}

 \maketitle

Upconversion is an optical process by which two optical fields
combine in a nonlinear medium to generate a third field at a
frequency equal to the sum of the two inputs
\cite{ref:Boyd_upconversion}. Along with energy conservation, this
process requires momentum conservation, placing a condition on the
phase velocities (or wavevectors) of the three beams.  Recently, it
has been shown that quasi-phase-matching \cite{ref:Fejer_IEEE}, a
process by which a grating in the nonlinear medium is used to
compensate for wavevector mismatch between the sum frequency beam
and the two input beams, can enable highly efficient upconversion
\cite{ref:Chanvillard}, with
near unity conversion efficiencies achieved in periodically-poled LiNbO$_3$ waveguides (PPLN WGs).  
Since upconversion, in principle, can be used to transduce one
photon at a given wavelength to a photon at another wavelength, one
could imagine using it to couple different quantum systems at
nominally different energies \cite{ref:Wallquist}. This kind of
hybrid quantum information scheme uses photons for efficient
transmission of quantum information over large distances
\cite{ref:Marcikic_Nat}, and robust, stationary quantum systems such
as trapped atoms \cite{ref:Boozer_Kimble} or ions
\cite{ref:Olmschenk_Monroe}, atomic ensembles
\cite{ref:Duan_Lukin_Cirac_Zoller,ref:Chaneliere_Kuzmich}, or spins
in quantum dots \cite{ref:Gerardot_Nat} for the manipulation and
storage of it. Another important application of quantum transduction
is the frequency upconversion of telecommunications-band photons to
the visible part of the spectrum for detection with
commercially-available low-noise silicon detectors, for example, as
part of a quantum information protocol. The challenges of single
photon detection in the near-infrared \cite{ref:Hadfield_nphoton}
inhibit the progress of linear optical quantum computation schemes
\cite{ref:Knill,ref:Kok_RMP} because of the stringent requirements
on efficient photon detection \cite{ref:Varnava_Rudolph}. That being
said, there have been a relatively limited number of upconversion
experiments in the quantum domain. In 1992, Huang and Kumar showed
\cite{ref:Huang_Kumar_PRL} that the non-classical intensity
correlation between twin beams at 1064 nm was preserved after one
beam was upconverted to 532 nm. In 2001, Kim, Kulik, and Shih used
\cite{ref:Kim_PRL} upconversion to implement a complete Bell state
measurement in a quantum teleportation scheme.  In 2005, Tanzilli
\textit{et al.} demonstrated \cite{ref:Tanzili_Zbinden} that the
time-energy entanglement between two modes at 1555~nm and 1312~nm
generated by spontaneous parametric downconversion was maintained
after the
1312~nm beam was upconverted to 712.4~nm; similar results have also been recently reported for a different set of wavelengths \cite{ref:Ma_entangle}.  
While non-classical correlations between the two beams were
demonstrated in each case after upconversion, taken individually,
each field exhibited classical photon statistics.  Recent progress
with upconversion detectors has allowed for photodetection at single
photon power levels \cite{ref:Vandevender_Kwiat_JMO,
ref:Langrock_Fejer,ref:Albota_Wong_upconversion, ref:Xu_Tang},
although in each case the light detected was a classical source (a
highly attenuated laser beam).  On the other hand, true
(non-classical) single photon generation has been shown to be
achievable through excitation of a single epitaxially-grown quantum
dot, with experimental demonstrations of stable sources in the 900
nm to 1300 nm region
\cite{ref:Michler,ref:Pelton,ref:Strauf_NPhot,ref:Zinoni_Fiore_SPAD,ref:Shields_NPhot},
but they have not been used for upconversion. Here, we demonstrate
for the first time, efficient upconversion of single photons
generated by a quantum dot at 1.3~$\mu$m to 710 nm. In particular,
we measure the second-order intensity correlation of the upconverted
710~nm signal and show that the field is dominantly composed of
single photons.
\\

\begin{figure*}
\centering
        \centerline{\includegraphics[width=13 cm,clip=true]{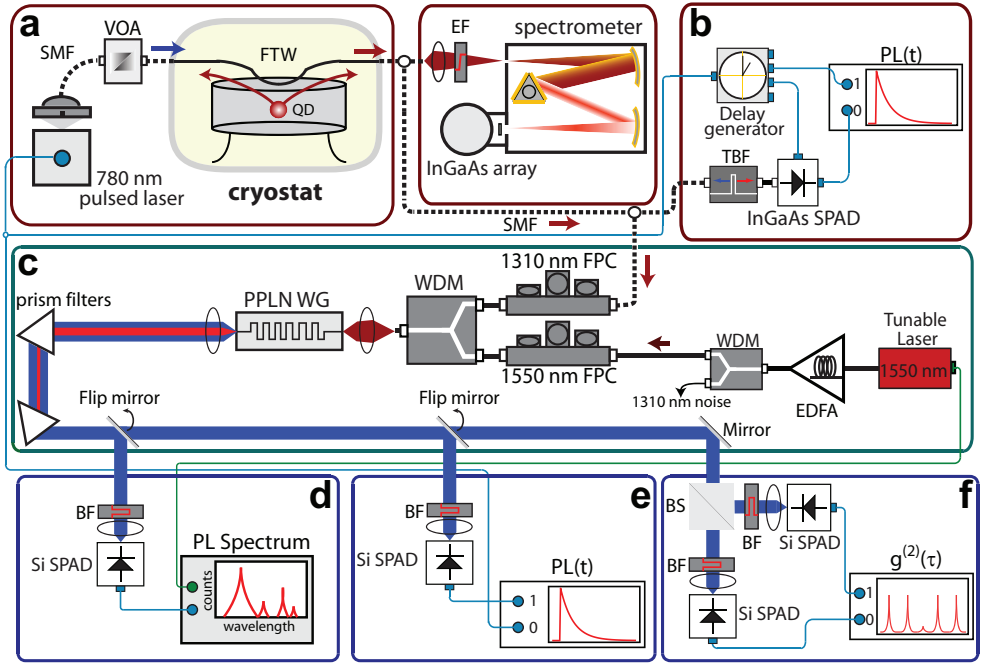}}
    \caption{\textbf{Experimental schematic for upconversion of photoluminescence from a QD.}  \textbf{a}, Schematic of experimental setup for excitation and collection of QD photoluminescence (PL) by a FTW.
    \textbf{b},  Schematic for time-resolved PL measurement using an
InGaAs SPAD.  \textbf{c}, Upconversion part of experimental setup.
\textbf{d}, Schematic for PL spectroscopy after upconversion.
    \textbf{e}, Schematic for time-resolved PL measurement after upconversion. \textbf{f}, Schematic of Hanbury-Brown and Twiss interferometer after upconversion.
    Definition of acronyms: PL=photoluminescence, SPAD=single photon counting
avalanche photodiode, FTW=fiber taper waveguide, SMF=single mode
fiber, VOA=variable optical attenuator, FPC=fiber polarization
controller, WDM=wavelength division multiplexer, EDFA= erbium-doped
fiber amplifier,
    TBF=tunable bandpass filter, EF=edgepass filter, BF=bandpass filter, BS=non-polarizing beamsplitter.}
    \label{fig:fig1}
\end{figure*}

\noindent \textbf{Optical Fiber-based Collection of Quantum Dot
Photoluminescence}

\begin{figure*}
\centering
        \centerline{\includegraphics[width=16 cm,clip=true]{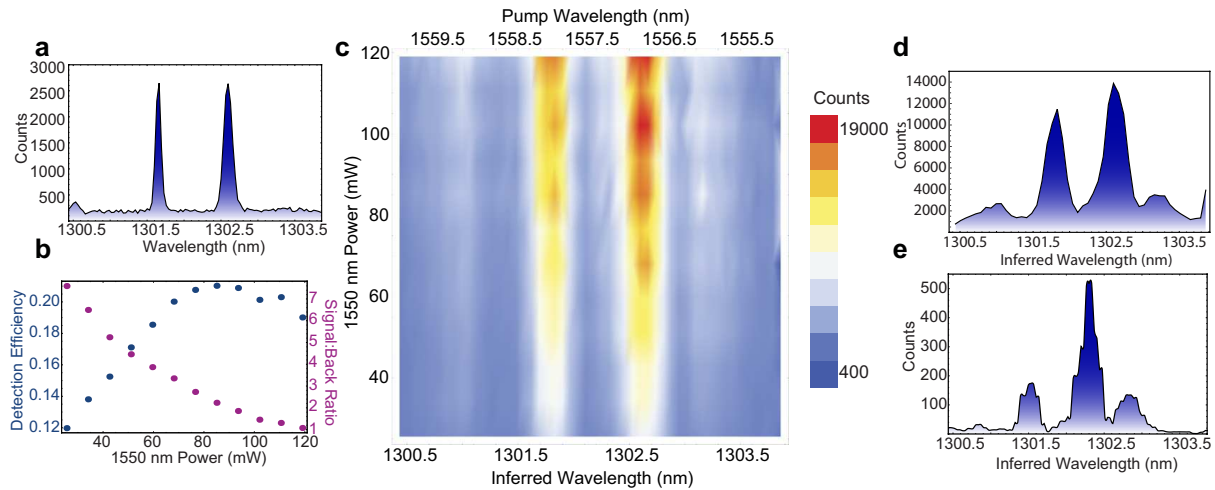}}
    \caption{\textbf{PL spectrum of a QD measured by upconversion.}  \textbf{a}, PL spectrum taken with a grating spectrometer with InGaAs array for 60 s
    integration time. \textbf{b}, Total detection efficiency (blue) and signal count to background count ratio (magenta) for upconversion detection
    as a function of 1550~nm pump power.
    \textbf{c},  Density plot of the upconverted PL spectrum taken by scanning the 1550 nm pump laser and integrating 1 s at each point as a function of 1550 nm pump laser power.
    \textbf{d}, Single upconverted PL spectrum from \textbf{c} with 85 mW of 1550~nm power. \textbf{e},  Upconversion response measured
    by recording spectrum of a laser near 1302.6~nm with 5 MHz linewidth.}
    \label{fig:fig2}
\end{figure*}

In our experiments, single photons at 1.3~$\mu$m are generated by
the recombination of excitons in a single epitaxially-grown InAs
quantum dot (QD) embedded in an InGaAs quantum well and within a 256
nm thick GaAs layer. This dot-in-a-well (DWELL) structure enables
efficient capture of excitons generated in the quantum well into the
QD\cite{ref:Stintz}. Following the work of
Ref.~\onlinecite{ref:Srinivasan15}, the QD-containing GaAs layer is
isolated 1.5~$\mu$m above the rest of the sample surface in a
2~$\mu$m diameter mesa (see Methods). The physical isolation enables
efficient probing of a limited number of QDs by a fiber taper
waveguide (FTW). The FTW is a single-mode optical fiber whose total
diameter is adiabatically reduced from 125 $\mu$m to roughly 1
$\mu$m over 10~mm by simultaneously pulling and heating the fiber
with a H$_2$ torch to near the melting temperature of glass
\cite{ref:Knight}. Because the diameter is comparable to the
wavelength of light passed along the fiber, the field exhibits a
non-negligible evanescent tail of a few hundred nm outside of the
physical dimension of the fiber, enabling efficient near-field
coupling to nanophotonic devices. As shown in the schematic of
Fig.~\ref{fig:fig1}a, the QD is optically excited using 50~ps pulses
from a 50 MHz repetition-rate gain-switched laser diode at 780 nm.
The laser is sent through a variable optical attenuator and reduced
to 10~nW of power (enough to saturate the single exciton QD
transition) before being coupled into the FTW by single mode fiber.
The FTW is incorporated into a continuous flow liquid Helium
cryostat and positioned into contact with the mesa, which is cooled
to a temperature of about 8 K.  The evanescent field of the pump
laser mode propagating through the FTW excites carriers into the
GaAs where they subsequently optically recombine in the QD,
generating photoluminescence (PL). A significant fraction of the
photons are emitted back into the FTW and coupled out into the fiber
(see Methods). Compared to free-space techniques using a lens or
high numerical aperture microscope objective, the FTW probe can
theoretically collect up to 10 times more photons \cite{ref:Davanco}
and provides a natural interface into single mode optical fiber for
convenient extraction without the need for alignment of optics. To
measure the QD emission spectrum, we couple the PL into a grating
spectrometer with a cooled InGaAs array with $\approx0.09$~nm
resolution as shown in Fig.~\ref{fig:fig1}a. Figure~\ref{fig:fig2}a
shows such a spectrum with 60 s integration time. The sharp lines of
a single QD are measured near 1301.6~nm and 1302.6~nm and identified
by polarization to be the positively charged and neutral exciton
transitions, respectively, of the same QD. For the QD shown here,
typical measured photon fluxes were $10^4$~s$^{-1}$, corresponding
to an average power of 1.5 fW.  At these low fluxes, InGaAs
detectors are difficult to use because of their high dark count
rates and lower gain which necessitate long integration times,
usually at least an order of magnitude longer than that required for
QDs with emission near 950~nm where Si-based detectors can be used.
\\

\noindent \textbf{Frequency Upconversion Spectroscopy}

For comparison, we also measure the PL spectrum using upconversion
as shown in Fig.~\ref{fig:fig1}c-d following the technique of
Ref.~\onlinecite{ref:Ma_upconversion},
\onlinecite{ref:Thew_upconversion}, and
\onlinecite{ref:Zhang_Fejer_upconversion}. In this experiment, the
PL is sent through a fiber polarization controller and combined with
a strong optical field near 1550 nm through a wavelength division
multiplexer. This field is obtained using a tunable external-cavity
diode laser coupled to an erbium-doped fiber amplifier (EDFA)
yielding 25 mW to 120 mW of continuous power into the PPLN WG. The
1550~nm light is sent through a fiber polarization controller before
being combined with the PL and directed into a 5~cm long,
temperature-controlled PPLN WG held at 61.0~$^{\circ}$C $\pm$
0.1~$^{\circ}$C. The strong optical nonlinearity of the PPLN WG
enables sum-frequency generation under appropriate
quasi-phase-matching conditions ($\approx0.35$~nm bandwidth) and
creates an optical field near 710 nm where the quantum efficiency of
a Si single photon counting avalanche photodiode (SPAD) is
$\approx$70~$\%$. The PPLN WG enables upconversion to around 710 nm
over a broad wavelength range of 1280 nm to 1340 nm by pump
wavelength and temperature adjustment. The light emerging from the
PPLN WG is dominantly composed of the upconverted 710~nm signal, the
remaining 1550~nm pump, and frequency-doubled pump photons at
775~nm. To remove the 1550~nm and 775~nm light, two dispersive
prisms and a 20 nm bandpass filter are employed before detection at
a Si SPAD. As shown in Fig.~\ref{fig:fig2}b, using a laser
attenuated to the same power level as the QD emission, we can
determine the total detection efficiency (blue) and signal to
background count ratio (magenta) as a function of 1550~nm input
power (see Methods). The overall detection efficiency maximum is
21.0~$\%$ $\pm$ 0.2~$\%$ for a pump power of 85 mW, while the signal
to dark count ratio is maximized for the minimum power of 25 mW.
Taking into account optical losses, a lower bound for the internal
conversion efficiency of the PPLN WG with 85 mW of 1550~nm power is
estimated to be 75~$\%$ $\pm$ 1~$\%$. To obtain a spectrum using
upconversion, the pump laser was spectrally tuned from 1555 nm to
1559.9 nm in 0.1 nm steps while pulses from the Si SPAD are counted
in 1~s bins; in contrast to Fig.~\ref{fig:fig2}a, a spectrum is
built-up through sequential single-channel measurements similar to
that of a scanning monochromator. The resulting spectra as a
function of 1550~nm pump power, pump wavelength, and inferred
1300~nm wavelength are shown in Fig.~\ref{fig:fig2}c, while
Fig.~\ref{fig:fig2}d shows a single spectrum taken with 85 mW of
1550 nm pump power. Note the presence of both QD emission lines as
in Fig.~\ref{fig:fig2}a. The spectrum is limited to a resolution of
$\approx$0.35~nm and contains side peaks because of the sinc$^2$
transfer function response of the PPLN WG \cite{ref:Fejer_IEEE}.
This instrument response can be measured directly using a narrow
linewidth (5 MHz) laser as shown for the spectrum in
Fig~\ref{fig:fig2}e.  Using the instrument response function it is
possible to deconvolve the measured spectrum, however, because the
QD transitions are much narrower than the spectral resolution,
deconvolution does not drastically improve the spectrum.  While this
is a limitation of the use of upconversion for high-resolution
spectroscopy, of more importance to subsequent photon counting
measurements is that a single-pixel signal-to-noise ratio similar to
that of Fig.~\ref{fig:fig2}a is obtained for 1/60 the integration
time.
\\

\noindent \textbf{Time-resolved Photoluminescence}

\begin{figure}
\centering
        \centerline{\includegraphics[width=8.5cm,clip=true]{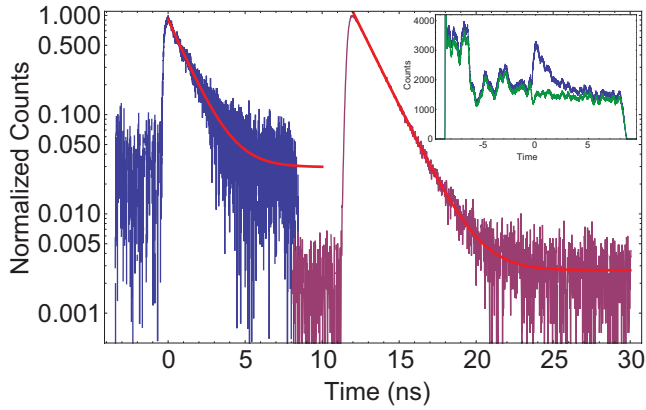}}
    \caption{\textbf{Time-resolved PL of a single QD.} PL lifetime measured by the InGaAs SPAD (blue) and the Si SPAD (maroon) after upconversion.
    Inset: Raw histograms of the signal (blue) and dark-count only (green) traces taken with the InGaAs SPAD.}
    \label{fig:fig3}
\end{figure}

An important counterpart to the aforementioned steady-state
spectroscopy is a time-resolved PL measurement which can reveal the
excited-state lifetime of the QD transition and provide insight into
effects such as radiative cascades and non-radiative decay. This
measurement can be obtained by incorporating a SPAD and a
time-correlated single photon-counting (TCSPC) board as shown
schematically in Fig.~\ref{fig:fig1}b. The TCSPC repeatedly records
the time interval between start events triggered by the pulsed laser
and stop events triggered by detection of a photon at the SPAD.
Then, the TCSPC generates a histogram of these time intervals which
corresponds to the time-evolution of the excited state of the QD
transition.  First, we used a commercial InGaAs SPAD to measure the
lifetime of the neutral exciton (1302.6 nm) peak after spectral
filtering by a tunable 1~nm bandpass filter.  Because of the
after-pulsing effects associated with InGaAs SPADs
\cite{ref:Ribordy}, the detector cannot run freely like a Si SPAD
and we must run in a gated detection mode as shown in
Fig.~\ref{fig:fig1}b. Our optimized detection settings were 20~$\%$
detection probability, 10~$\mu$s dead time, and a 20 ns gate width
triggered by a delay generator with delay chosen to match the
detection window to the arrival of pulses of PL. Under this
configuration, the trigger rate was 4.3~MHz and the measured dark
count rate was 12~500~s$^{-1}$. Because of quantum efficiency
oscillations that occur at the start of the gate of our InGaAs
detectors, we chose a sufficiently long gate width to avoid these
modulations and include the full decay of the PL
\cite{ref:Zinoni_Fiore_SPAD}.  The lifetime trace after background
subtraction is shown in blue in Fig.~\ref{fig:fig3} for an
integration of 700~s. The raw data for the signal (blue) and dark
count only (green) measurements are shown in the inset of
Fig.~\ref{fig:fig3} where the quantum efficiency oscillations are
clearly visible along with the arrival of the optical pulse at 0~ns.
Analysis of the background-subtracted trace leads to a measured
lifetime (with 95~$\%$ confidence interval) of 1.3~ns$\pm$ 0.1~ns
and the resulting fit is shown along with the data.

The same lifetime measurement can be performed using a Si SPAD after
upconversion as shown in Fig.~\ref{fig:fig1}e.  The PL is combined
with 25.5 mW of the pump laser at 1556.8~nm in the PPLN WG for
efficient upconversion of the neutral exciton transition at
1302.6~nm to 710~nm.  Because of the 0.35~nm upconversion bandwidth,
no additional spectral filtering is required to select the neutral
exciton peak. The freely running SPAD (50~ns dead time, 100~s$^{-1}$
intrinsic dark count rate) provides the stop signal for TCSPC and we
integrate for 600~s. To remove background counts, a measurement
without the 780~nm excitation is also taken for subsequent
subtraction. The resulting measurement in shown in maroon in
Fig.~\ref{fig:fig3} and has been temporally offset for comparison
with the InGaAs measurement.  A fit to the data (red line) yields a
lifetime (with 95~$\%$ confidence interval) of 1.38~ns$\pm$ 0.03~ns.
Figure~\ref{fig:fig3} clearly shows the Si measurement has a dynamic
range $\approx$25 times better than that of the InGaAs SPAD and
demonstrates the usefulness of upconversion for substantially more
sensitive detection.
\\

\noindent \textbf{Second-order Intensity Correlation}

\begin{figure}
\centering
        \centerline{\includegraphics[width=8.5cm,clip=true]{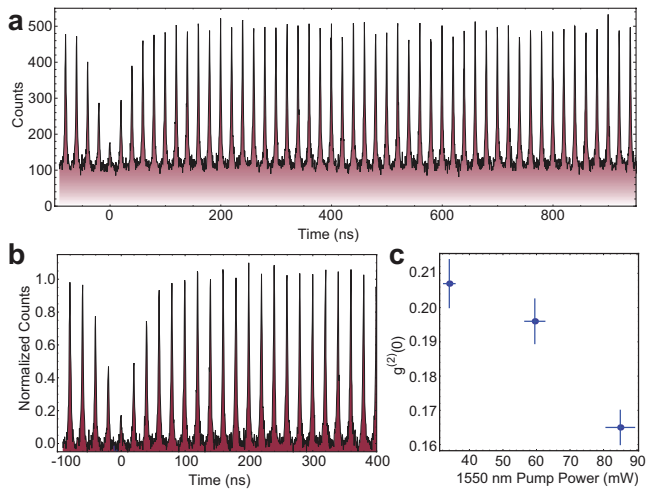}}
    \caption{\textbf{Upconversion of single photons.}  \textbf{a}, Raw histogram of the second-order intensity correlation measurement after
    3600~s. \textbf{b}, Normalized second-order intensity correlation
    $g^{(2)}(\tau)$ after ASR photon event subtraction. \textbf{c},
    $g^{(2)}(0)$ values with errors for different powers of the 1550~nm pump laser.}
    \label{fig:fig4}
\end{figure}

To prove that the upconverted 710 nm signal was composed of single
photons and inherently non-classical, we measured the second-order
intensity correlation $g^{(2)}(\tau) = \langle:
I(t)I(t+\tau):\rangle/\langle I(t)\rangle^2$ using a Hanbury-Brown
and Twiss interferometer as depicted in Fig.~\ref{fig:fig1}f.  After
upconversion in the PPLN WG pumped by 85~mW of 1556.8~nm power and
spectral dispersion by the prisms, the 710~nm signal was divided by
a 50:50 non-polarizing beamsplitter and then directed into two Si
SPADs.  The electronic pulses by the SPADs were used as the start
and stop signals for the TCSPC.  The histogram of such start-stop
events as a function of the time interval is directly related to
$g^{(2)}(\tau)$ and is shown in Fig.~\ref{fig:fig4}a.  The data
shows that the areas between PL pulses every 20~ns have a non-zero
number of events. Clearly, this time-independent background arises
from start-stop events triggered by detection of photons not from
the PL of the QD, which decays on a fast timescale ($\tau_{PL} =
1.38$~ns as in Fig.~\ref{fig:fig3}) every 20~ns.  The origin of
these events is the start and/or stop detection of upconverted
anti-Stokes Raman (ASR) photons near 1302.6~nm from the pump laser
at 1556.8~nm that are transmitted by the 20~nm bandpass filter and
compose approximately 1/3 of the total counts (see
Fig.~\ref{fig:fig2}b). Nonetheless, the peak at $\tau=0$ is reduced
compared to the other peaks by a factor of 0.37$\pm$0.02 $<0.5$,
indicating the single photon nature of the light.  However, because
the ASR photon events are temporally distinguishable from true PL
events, this background can be subtracted (see Methods) to obtain
the true $g^{(2)}(\tau)$ for the PL signal. In principle, the
background can be easily reduced by using smaller bandwidth filters;
for instance a 1~nm filter would reduce it by approximately a factor
of 20. Another feature of Fig.~\ref{fig:fig4}a is that the peaks
nearest $\tau=0$ at $\tau = \{\pm 20$~ns, $\pm 40$~ns$\}$ do not
recover completely even though the QD lifetime is only 1.38~ns. This
sub-microsecond correlation effect has been measured previously for
QDs \cite{ref:Santori4} and is thought to be caused by the
preferential capture of single carriers into the QD rather than
electron-hole pairs. The strength of this effect is known to depend
strongly on excitation wavelength and power and is not the focus of
this work. After ASR background subtraction and normalization by
peaks far from $\tau=0$, $g^{(2)}(\tau)$ can be extracted as shown
in Fig.~\ref{fig:fig4}b. The reduction at $\tau=0$ is a clear
indication of photon anti-bunching and the non-classical nature of
the optical field. The value $g^{(2)}(0)$ can be obtained by
comparing counts in the $\tau=0$ peak to the average of those in the
$\tau \neq 0$ peaks and we obtain a value of $g^{(2)}(0)=0.165$
$\pm$ $0.010$.  Because $g^{(2)}(0)=0.165 < 0.5$, the optical field
must be dominantly composed of single photons. The measured non-zero
value of $g^{(2)}(0)=0.165$ is due to uncorrelated emission at 1.3
$\mu$m collected by the FTW and subsequently upconverted. The number
of uncorrelated photons may be reduced by exciting closer to the QD
transition.  In addition, the 1550~nm pump power might also be
expected to influence the $g^{(2)}(0)$ value and the signal-to-noise
given tradeoffs between increased detection efficiency and higher
levels of background ASR photons as shown in Fig.~\ref{fig:fig2}b.
To that end, we performed the $g^{(2)}(\tau)$ measurement at three
values of the 1550~nm pump power and the results are displayed in
Fig.~\ref{fig:fig4}c.  The best result (Fig.~\ref{fig:fig4}b) was
obtained for 85 mW of power which corresponded to the optimum
overall detection efficiency (21.0~$\%$) of our setup.
\\

\noindent \textbf{Summary}

In conclusion, we have used frequency upconversion in a PPLN WG to
efficiently measure single photon PL from a QD emitting near 1.3
$\mu$m with Si SPADs.  In doing so, we have demonstrated that
upconversion can be used for high signal to noise spectrometry and
time-correlated photon counting for lifetime measurements.  While
other groups have shown that superconducting single-photon detectors
can be used instead of InGaAs SPADs for such measurements in the
near-infrared \cite{ref:Zinoni_Fiore_SSPD}, those detectors are not
yet widely available and require cryogenic cooling whereas efficient
upconversion can be performed with off-the-shelf components.
Additionally, we showed that the quantum mechanical nature of the
single-photon stream emitted by the QD at 1.3 $\mu$m is successfully
transduced over to a 710~nm optical field and in particular we
measure $g^{(2)}(0)=0.165$ for this field.  Aside from the clear
advantage offered in increased measurement sensitivity, efficient
quantum transduction by frequency upconversion has several
consequences for quantum information schemes involving stationary
quantum systems and photons.  Because of the large upconversion
wavelength range afforded by the quasi-phase-matching of the PPLN
WG, QDs with different emission wavelengths as large as a few nm (to
be expected due to QD growth inhomogeneity) can all be upconverted
to the same wavelength. A clear implication of this is that remote
operations in the telecommunications-band such as the generation of
entanglement \cite{ref:Simon} between spatially separated electron
spins in QDs is possible without the fine spectral-tuning typically
required in such schemes. Further, one can envision photon-mediated
interactions between electron or hole spins in QDs in the
near-infrared with alkali atoms or ions in the near-visible. For
example, a hole confined in a QD with an optical transition at
1.3~$\mu$m could be coupled to a Rb atom or ensemble
\cite{ref:Eisaman_Lukin} at 780~nm by upconversion with a
thulium-doped fiber laser at 1.95~$\mu$m. Finally, because the
upconversion process depends critically on the strong pump field,
fast temporal sampling of the input field is possible by using
ultrafast pump pulses \cite{ref:kuzucu_wong,ref:Shah_JQE}. At
present, many time-dependent measurements of QD PL are at best
limited to $\approx$40~ps resolution by the Si SPAD timing jitter.
Given that pulse widths as short as 4.3~fs have been created using
mode-locked fiber lasers \cite{ref:Krauss_NatP}, QD dynamics on
ultrafast timescales could be revealed by upconversion.

\noindent\textbf{Methods}

\noindent\textbf{Mesa Fabrication:} The QD-containing mesas are
created through: (i) electron beam lithography, (ii)
SF$_6$/C$_4$F$_8$ inductively-coupled plasma reactive ion etching
(ICP-RIE) of a deposited Si$_{x}$N$_{y}$ mask layer, (iii) Ar-Cl$_2$
ICP-RIE etching of the Al$_{0.3}$Ga$_{0.7}$As layer and removal of
the remaining Si$_{x}$N$_{y}$ layer, and (iv) 20:1 H$_2$O:HF wet
etching of the underlying Al$_{0.7}$Ga$_{0.3}$As layer to form the
supporting pedestal.

\noindent\textbf{FTW PL collection efficiency:} The FTW PL
collection efficiency is determined from the measured Si SPAD count
rates corrected for the total detection efficiency (21~$\%$) and
then normalized to the 50 MHz excitation repetition rate.  The
normalization is the maximum rate of photon emission by the QD
assuming it is excited a single time for each pump pulse.  This
assumption is valid because the pulse width (50~ps) is much shorter
than the radiative lifetime (1.38~ns) and the average power is such
that the transition is saturated. For this QD which resides in a
non-undercut portion of the mesa, we obtain an efficiency near
$0.1~\%$. This efficiency is consistent with previous experimental
work \cite{ref:Srinivasan15} and matches theoretical predictions
\cite{ref:Davanco} for a non-undercut dielectric substrate.

\noindent\textbf{Detection and conversion efficiencies:} The overall
detection efficiency of $21~\%$ is determined with a laser
attenuated to a measured power equivalent to the QD (1.5 fW) and set
at the wavelength of the neutral exciton transition. The measured Si
SPAD count rates after background subtraction are compared to the
known input power level. The internal conversion efficiency of
$75~\%$ is calculated by correcting the overall efficiency by the
measured optical losses ($\eta_{WDM}=0.95$, incoupling and
transmission through the PPLN WG $\eta_{PPLN}=0.61$, beamsplitter
and mirror loss $\eta_{BS}= 0.81$, and $\eta_{BF}=0.85$) and Si SPAD
quantum efficiency at 710 nm ($\eta_{SPAD}=0.70$).  The errors in
both efficiencies dominantly result from shot noise in the photon
counts. Including the PL collection efficiency, the end-to-end
efficiency of the system is $\approx0.02~\%$.

\noindent\textbf{$\mathbf{g^{(2)}(\tau)}$ background subtraction:}
As described in the text, background counts are due to ASR photon
events and are time-independent.  The subtracted level was
determined by taking the average number of counts measured for
time-delays greater than 8.2~ns from the nearest PL pulse.  Since
the lifetime was measured to be 1.38~ns, the probability of a PL
event contributing at these time delays is negligible
($\approx0.3~\%$).  The background counts observed in the
measurement are consistent with what is expected from the ratio of
signal to background counts measured in Fig.~\ref{fig:fig2}b for
85~mW of pump power.  Errors in the extracted $g^{(2)}(0)$ values
are determined by the propagation of errors due to shot noise in the
$\tau=0$ peak area, the standard deviation of the peak areas far
from $\tau=0$, and the shot noise in the subtracted background
level.

\noindent \textbf{Acknowledgements} We thank A. Stintz and S.
Krishna of the University of New Mexico and O. Painter of the
California Institute of Technology for assistance with sample
preparation, and M. Davan\c{c}o of NIST for development of the fiber
taper fabrication setup.

\noindent \textbf{Author Contributions} M.T.R. and K.S. built the
low-temperature measurement setup, L.M., O.S., and X.T. built the
upconversion detectors, K.S. fabricated the devices, and M.T.R.,
L.M., and K.S. performed the measurements. M.T.R. and K.S. wrote the
manuscript, all authors contributed to the design of the
experiments, and K.S. and X.T. supervised the project.

\noindent \textbf{Author Information} The authors declare no
competing financial interests.  Correspondence and requests for
material should be address to K.S.

\end{document}